# GROWTH, STRUCTURE AND PROPERTIES OF EPITAXIAL THIN FILMS OF FIRST PRINCIPLES PREDICTED MULTIFERROIC Bi$_2$FeCrO$_6$

Riad Nechache, Catalin Harnagea, and Alain Pignolet[1]

*INRS Énergie, Matériaux et Télécommunications,*

*1650, Boulevard Lionel Boulet, Varennes, Québec, Canada, J3X 1S2*

François Normandin, and Teodor Veres

*Industrial Materials Institute, National Research Council Canada,*

*75, de Mortagne, Boucherville, Québec, Canada, J4B 6Y4*

Louis-Philippe Carignan and David Ménard

*Department of Engineering Physics, École Polytechnique de Montréal,*

*P.O. Box 6079, Station. Centre-ville, Montréal, Québec, Canada, H3C 3A7*

---

[1] Electronic-mail : pignolet@emt.inrs.ca

**Abstract**

We report the structural and physical properties of epitaxial $Bi_2FeCrO_6$ thin films on epitaxial $SrRuO_3$ grown on (100)-oriented $SrTiO_3$ substrates by pulsed laser ablation. The 300 nm thick films exhibit both ferroelectricity and magnetism at room temperature with a maximum dielectric polarization of 2.8 µC/cm$^2$ at $E_{max}$ = 82 kV/cm and a saturated magnetization of 20 emu/cc (corresponding to ~ 0.26 µ$_B$ per rhombohedral unit cell), with coercive fields below 100 Oe. Our results confirm the predictions made using ab-initio calculations about the existence of multiferroic properties in $Bi_2FeCrO_6$.

Multiferroics are materials which display both ferroelectric and magnetic ordering and are promising candidates for innovative devices such as transducers, sensors[1] and memories for data storage. Recently, using first-principles density functional theory, Spaldin *et al.*[2] studied $Bi_2FeCrO_6$ (BFCO) and predicted ferrimagnetic and ferroelectric properties allegedly much larger than those of the known multiferroic materials. They theoretically predicted that at 0 K BFCO would be both ferrimagnetic and ferroelectric with a magnetic moment of $2\mu_B$ per formula unit and a polarization of 80 µC/cm$^2$, respectively. The predicted ground state structure is very similar to that of the rhombohedral R3c structure of $BiFeO_3$ (BFO), but with $Fe^{3+}$ cations substituted with $Cr^{3+}$ cations in every second (111) layer, reducing the symmetry to the space group R3.

To our knowledge, the successful synthesis and the measurement of structural and multiferroic properties of thin films of this hypothetical material have never been reported. Here are presented the successful synthesis of $Bi_2FeCrO_6$ epitaxial films by pulsed laser deposition (PLD) and their structural, electric and magnetic characterization. These results are compared to the properties predicted by *ab-initio* calculations.[3]

Pulsed laser deposition (PLD) is one of the most suitable and frequently used techniques to grow heterostructures of complex multi-component oxides in moderate oxygen pressure. Using this process we prepared heterostructures of BFCO / $SrRuO_3$ films on (100) oriented $SrTiO_3$ (STO) substrates. The epitaxial $SrRuO_3$ (SRO) film was prepared from a stoichiometric ceramic target and was used both as bottom electrode for electrical measurements and to promote the heteroepitaxial growth of BFCO. The epitaxial BFCO films were obtained from a dense ceramic target composed of a stoichiometric mixture of $BiFeO_3$ and $BiCrO_3$. The substrate temperature during deposition was 750°C and the BFCO films were deposited at a growth rate of ~1.5 Å/sec in oxygen ambient at a pressure of 20 mTorr. To eliminate possible oxygen vacancies, the films were cooled from 750°C down to 400°C at a slow cooling rate of 8°C/min in 1 atm of oxygen, and maintained at this temperature for one hour, prior cooling down to room temperature.

The $Bi_2FeCrO_6$ film stoichiometry was investigated using energy dispersive x-ray spectroscopy (EDX), X-ray photoelectron spectroscopy (XPS), as well as Rutherford Backscattering (RBS). XPS depth profiling analyses and EDX in TEM reveal a composition uniform throughout the film thickness and the three techniques agree on the correct surface cationic stoichiometry (Bi:Fe:Cr = 2:1:1). The crystal structure of the BFCO films was investigated using x-ray diffraction (XRD) (PANalytical X'Pert MRD 4-circle diffractometer). The BFCO film is epitaxial and we did not observe any reflections that would be indicative of any secondary phase. In Figure 1a a detail of the θ/2θ scan at high angle (66° - 76°) is presented showing the 003 peaks of BFCO near the 300 reflection of STO. A XRD spectrum of an epitaxial $BiFeO_3$ (BFO) film is shown as well. The out-of-plane lattice parameters were calculated to be 3.98 Å for BFO and 3.97 Å for BFCO. Comparing the XRD spectra of BFCO and BFO for 300 nm thick epitaxial films, we find that they are similar, reflecting an equivalent crystal structure in agreement with theoretical prediction.[3] In both spectra, the 003 reflection of the 60 nm thin SRO layer is not visible since it is located at an angle close to that of BFO, respectively of BFCO, and thus is buried within the 003 reflection of the upper layer. Such a high value of the out-of-plane lattice parameter of the 60 nm thick epitaxial SRO bottom electrode is indicative of a fully strained epitaxial SRO layer. The in-plane orientation was investigated by Φ-scan, using the pseudo-cubic {211} reflections of both BFCO and STO (figure 1b). The fourfold symmetry and relative position of the STO and BFO peaks indicate a "cube-on-cube" epitaxy of BFCO on STO/SRO. Selected area electron diffraction (SAED) patterns (figure 1c), as well as low-magnification bright field TEM images (not shown) obtained from a cross-section of a (001)-oriented BFCO film confirm the high crystalline quality of the BFCO film a and did not provide any evidence of second phases or inclusions of any kind. The c-axis of BFCO is found to be parallel to the film normal and only 001 and 210 reflections from a pseudo-cubic phase are observed. A further analysis of the SAED pattern reveals that the BFCO layer has an out-of-plane lattice parameter of 3.95 Å and an in-plane lattice parameter very close to that of the STO substrate (~3.91 Å). These lattice parameters are 2%, respectively 1%, larger than the ones predicted by first-principles density functional theory.[3]

Ferroelectric properties were investigated macroscopically using a commercially available Thin Film Analyzer (TFA2000, aixACCT Systems GmbH). Pt electrode dots were deposited through a mask, by pulsed laser ablation (PLD) on the top of the BFCO layer. A ferroelectric hysteresis loop (figure 2a), measured between a Pt top electrode and the SRO bottom electrode at 100Hz, reveals both a ferroelectric and a leaky behavior of the BFCO films. The maximum polarization along the [001] direction of the pseudo-cubic unit cell measured at an applied electric field of 82 kV/cm reaches 2.8 $\mu C/cm^2$. Fully saturated hysteresis loop could not be obtained due to the leaky nature of the sample, as shown by the I-V characteristic in the inset of figure 2a, where we measured a current density of 800 $\mu A/cm^2$ at a dc electric field of 100 kV/cm. The dielectric properties were also measured using the same equipment. The effective relative dielectric permittivity (measured at 1 kHz) in a 300 nm BFCO layer was found to be around 275 with a loss tangent of 0.8 for an electric field of 100 kV/cm.

Local piezoelectric measurements were carried out using piezoresponse force microscopy (PFM)[4,5,6] alleviating the problems related to leakage currents. Here we used a DI-Enviroscope AFM (Veeco) equipped with a NSC36a (Micromasch) cantilever and tips coated with Co/Cr. We applied an ac voltage of 0.5V at 26 kHz between the conductive tip and the SRO layer of the sample located beneath BFCO and we detected the BFCO film surface induced piezoelectric vibrations using a Lock-in Amplifier from Signal Recovery (model 7265). Using a dc-source we alternately applied voltages of minus and plus 14V to the bottom electrode during the scanning in contact mode. The resulting contrast shown in the inset of figure 2b is a clear proof that ferroelectricity in BFCO exist and can be switched upon application of an external voltage. A further proof of ferroelectricity, is the presence of a piezoelectric hysteresis loop obtained with the AFM-tip fixed above the location of the sample surface being tested (remanent loop in this case).[4] The strength of the PFM signal is comparable to that obtained from BFO films.

Since the *Ab-initio* calculation results predict ferrimagnetic magnetization, magnetic properties of the films were estimated using a Quantum Design Physical Property Measurement System (PPMS) and the DC extraction method with a sensitivity of $2 \times 10^{-5}$ emu. The measurements were verified at room temperature using a Vibrating Sample Magnetometer (VSM) with a better sensitivity of $10^{-6}$ emu. The room-temperature magnetic hysteresis loop of a 300 nm thick BFCO film is compared to that of a BFO film having the same thickness in figure 3. The magnetic field H was applied in the plane of the films, parallel to the [100] direction of the substrate. The saturated magnetization of BFCO was found to be ~20 emu/cc at room temperature, corresponding to ~0.26 $\mu_B$ per two 'formula unit' rhombohedral unit cell. This room temperature value is lower than the value of 2 $\mu_B$ per Fe–Cr pair predicted by *ab-initio* calculation for bulk BFCO at 0K, which is to be expected. These magnetic measurements clearly show a hysteretic behavior of the BFCO films with a coercive field of 80 Oe, which is about half of that measured for BFO films. Nonetheless, when comparing the room temperature magnetic properties of BFCO to those of a BFO film of the same thickness, we do notice a ten-fold increase of the saturation magnetization of the BCFO film. The remnant magnetization of BFCO (2$M_r$ = 2 emu/cc), however, is showing only a four-fold increase. Note that for our epitaxial BFO layers, the saturation magnetization measured was ~ 4 emu/cc per two 'formula unit' rhombohedral unit cell (or ~ 2 emu/cc per pseudo-cubic unit cell), which is consistent with the values previously reported for pure BFO thin films of similar thicknesses.[7-11] Preliminary results also indicate an increase of magnetization with decreasing thickness for both BFCO and BFO films, as reported by Wang *et al.* for BFO.[7,12] The presence of magnetization hysteresis at room temperature suggests that the magnetic ordering temperature exceeds the theoretically predicted one[3] (110 K) and even surpasses the value found very recently by Yu and Ithoh in bulk BFCO ceramics, which has been extrapolated to be around ~220K.[13]

In summary, we successfully prepared epitaxial BFCO films by pulsed laser deposition. The crystal structure found is very similar to that of BFO, and the films have the correct cationic stoichiometry throughout their thickness. The BFCO films are leaky but exhibit good ferroelectric and piezoelectric properties at room temperature. The insulating properties of the films are not yet optimized, possibly due to a slight loss of Bi during deposition. Magnetic measurements show that the films exhibit a magnetization hysteresis at room temperature with a saturation magnetization about one order of magnitude higher than that of BFO films having the same thickness. Our results partly confirm the predictions made using the *ab-initio* calculations about the existence of multiferroic properties in BFCO. The existence of magnetism at room temperature is a very promising, yet unexpected result that needs to be further investigated, and studies of the magnetic ordering and of the magnetoelectric coupling in the films are currently underway.

**Acknowledgments**

The authors want to thank Dr. Philippe Plamondon ((CM)$^2$, École Polytechnique de Montréal) for Xray analysis in TEM and SAED analysis and related discussions. Part of this research was supported by INRS start-up funds, NRC-IMI operational funds, NSERC (Canada), and FQRNT (Québec).

**FIGURE CAPTIONS**

**Figure 1** (**a**) X-ray θ/2θ scan of the 003 pseudocubic reflection for a 300 nm thick epitaxial BFCO film (black) and for a BFO (blue) layer of the same thickness, both deposited on (100)-oriented STO substrates coated with an epitaxial 60 nm thick SRO(100) conductive layer. (**b**) Φ-scan using the 211 reflection of BFCO and the 211 reflection of STO from the same BFCO/SRO/STO sample. (**c**) Selected area electron diffraction pattern of the heterostructure taken along the [1 $\bar{2}$ 0] beam direction. Reflections 001, 210 and 211 are indexed according to the pseudo-cubic unit cell of BFCO.

**Figure 2** Ferroelectric properties of a 300 nm thick BFCO film, **(a)** Ferroelectric hysteresis loop at room temperature, **(b)** Local remanent piezoresponse hysteresis. The inset shows a PFM image (scan size 10 X 10μm) after writing oppositely polarized domains using the AFM tip as a top electrode (black/white contrast represents polarization oriented upwards/downwards)

**Figure 3** Magnetization hysteresis of a 300 nm thick BFCO film (squares) compared to that of a 300 nm thick BFO film (circles) at room temperature. The inset shows a zoom around the origin. The direction of applied magnetic field H was in the plane of the films, along the [100] direction of STO.

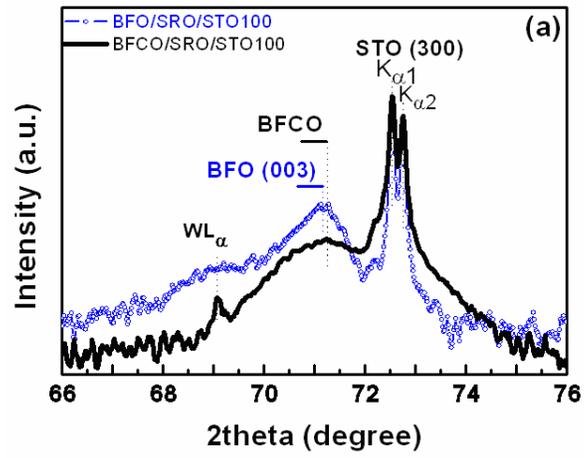
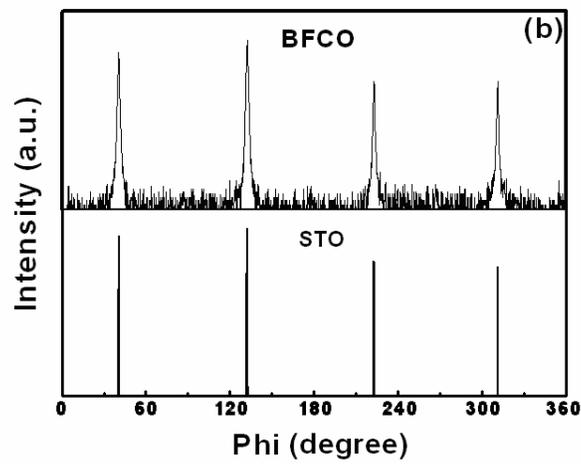
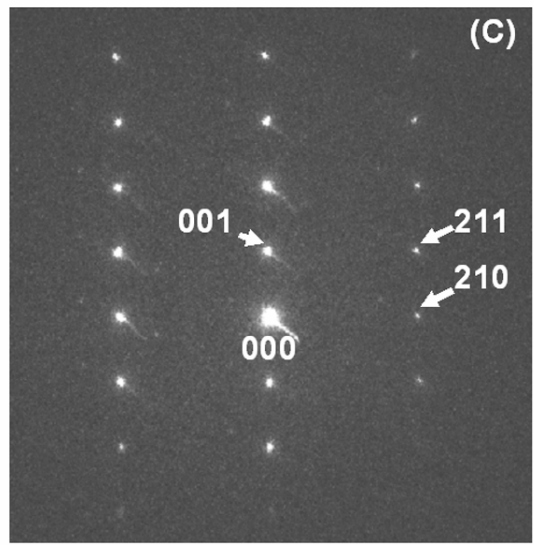

**Figure 1, R. Nechache et al.**

**Submitted to Applied Physics letters**

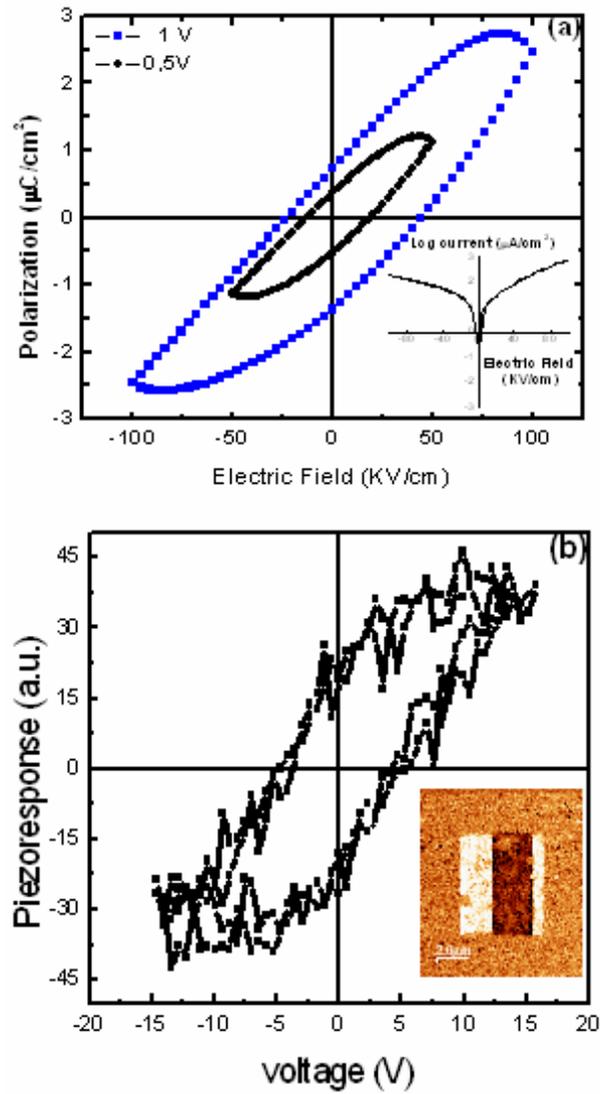

**Figure 2, R. Nechache et al.**

**Submitted to Applied Physics letters**

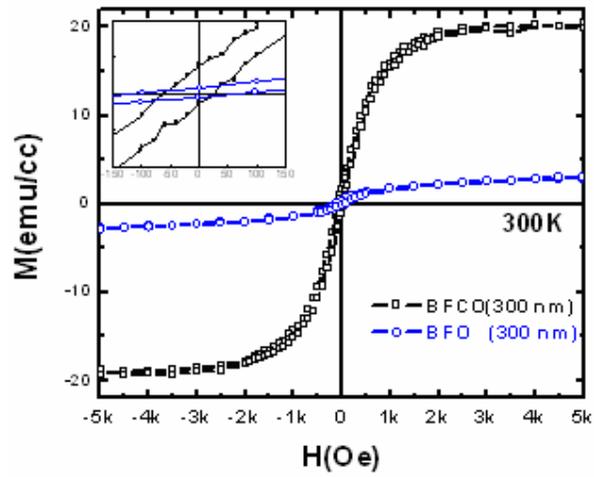

**Figure 3, R. Nechache et al.**

**Submitted to Applied Physics letters**